# Long dephasing time and high temperature ballistic transport in an InGaAs open quantum dot

B. Hackens[a,*], S. Faniel[a], F. Delfosse[a], C. Gustin[a], H. Boutry[a], I. Huynen[a], X. Wallaert[b], S. Bollaert[b], A. Cappy[b], V. Bayot[a]

[a]*CERMIN, PCPM, DICE and EMIC Labs, Universite Catholique de Louvain, Louvain-la-Neuve, Belgium*

[b]*IEMN, Cite scientifique, Villeneuve d Ascq, France*

[*] *Corresponding author. Tel.: +32-10-478191; fax: +32-10-473452; e-mail: hackens@pcpm.ucl.ac.be.*

**Abstract**

We report on measurements of the magnetoconductance of an open circular InGaAs quantum dot between 1.3K and 204K. We observe two types of magnetoconductance fluctuations: universal conductance fluctuations (UCFs), and focusing fluctuations related to ballistic trajectories between openings. The electron phase coherence time extracted from UCFs amplitude is larger than in GaAs/AlGaAs quantum dots and follows a similar temperature dependence (between $T^{-1}$ and $T^{-2}$). Below 150K, the characteristic length associated with focusing fluctuations shows a slightly different temperature dependence from that of the conductivity.

## 1. Introduction

Transport in semiconductor quantum dots (QDs) is mainly determined by two characteristic times: the phase coherence time $\tau_\phi$ and the elastic relaxation time $\tau_e$. The phase coherence time is the key parameter which determines whether a mesoscopic system either behaves in a classical way or exhibits quantum mechanical effects, such as electron interferences. On the other hand, the ballistic transport regime is reached when $\tau_e$ becomes comparable to the electron transit time through the QD. The ability to control and to increase both timescales is essential in view of potential use in *e.g.* quantum information processing or high frequency applications [1].

Here, we extract $\tau_\phi$ from the magnetoconductance of an InGaAs open quantum dot, and find larger values than previously reported in AlGaAs/GaAs quantum dots [2]. We also concentrate on the signatures of ballistic electron behavior inside the dot. We study their temperature dependence and notice their persistence up to unexpectedly high temperatures.

**2. Experiment**

The starting material for our sample is an InGaAs/InAlAs heterostructure, with an electron density $n_s = 10^{16}$ $m^{-2}$ and a mobility ~3 $m^2$/Vs at 4K (equivalent to an elastic mean free path $l_\mu$~ 500 nm). Using electron beam lithography and wet etching, we define a circular quantum dot, with a lithographic diameter of 430 nm and openings width of 60 nm. The depletion layer at the edges of the structure is approximately 25 nm wide. A metallic gate deposited on the whole structure allows us to tune its shape and change the electron density. The conductance of the sample is measured using a four-contacts lock-in technique. The measurements are performed at temperatures between 1K and 204 K and in a magnetic field *B* up to 5T.

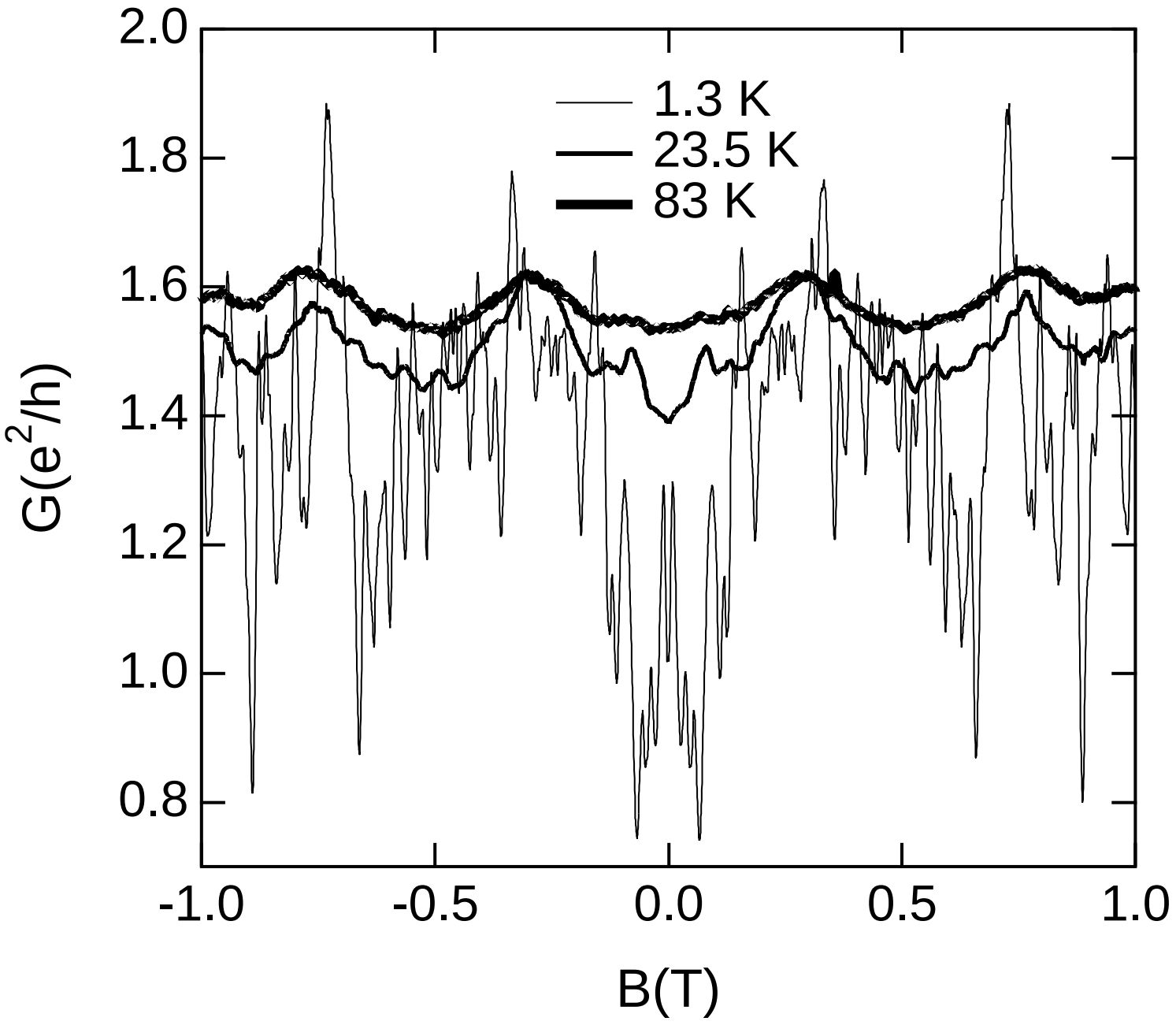

Fig. 1. Temperature dependence of the magnetoconductance in units of $e^2$/h.

**3. Results**

Fig. 1. shows the temperature dependence of the magnetoconductance *G(B),* with the gate grounded. The average conductance of the cavity is ~ 1.4 $e^2$/h to 1.6 $e^2$/h, which means that there is at least one transverse mode

in the openings. The magnetoconductance exhibits two types of fluctuations, each with a different temperature dependence. The first ones, also called universal conductance fluctuations (UCFs), are related to quantum interference phenomena inside the dot and persist up to 30K. They are superimposed on a second type of fluctuations, which occur on a much wider magnetic field range, and are more robust in temperature (observed up to 204K). Most of these wide maxima and minima can be associated with direct focusing trajectories between entrance and exit point contacts. Both types of fluctuations are symmetric with respect to $B = 0$.

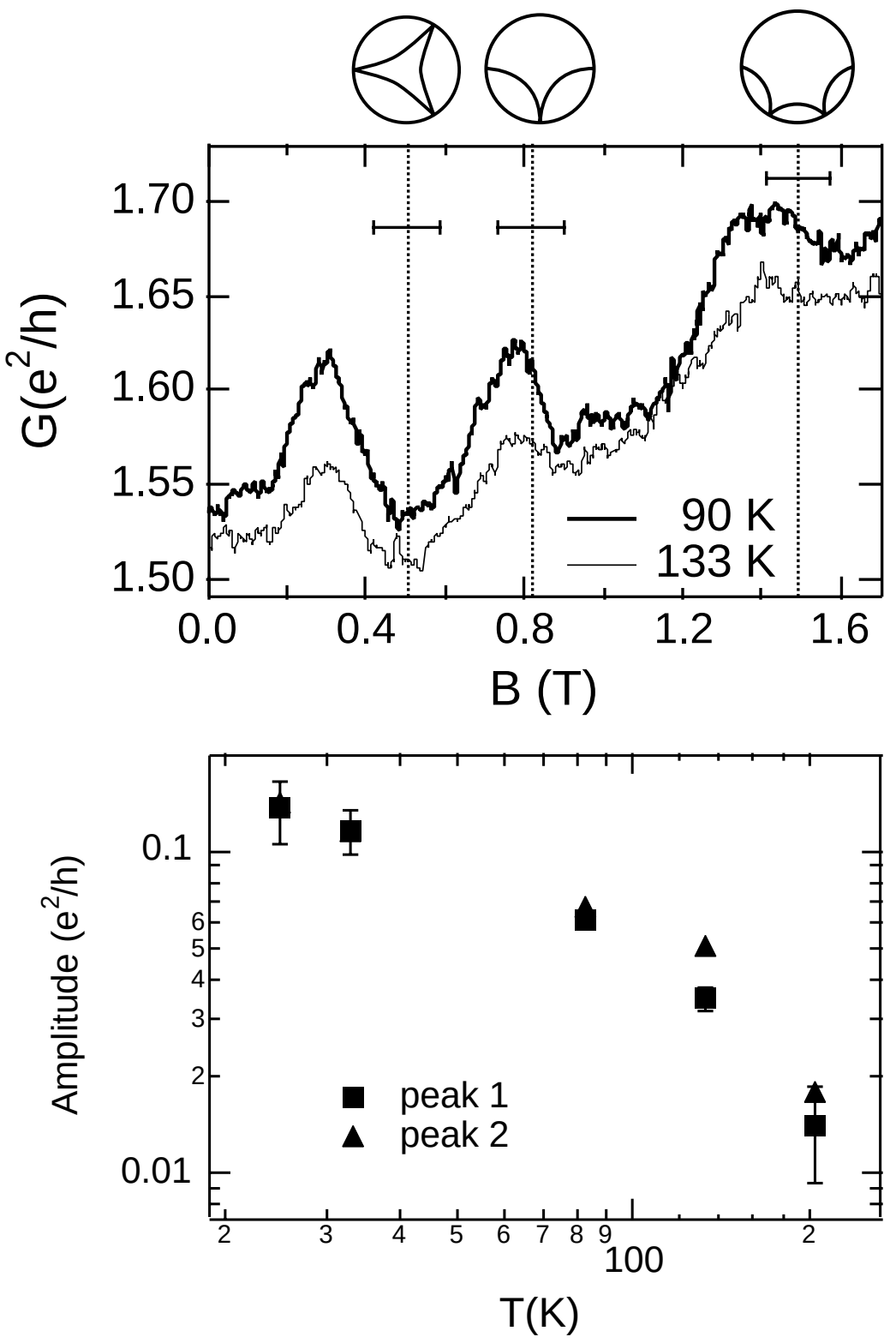

Fig. 2. Top: high temperature magnetoconductance, with the magnetic field position of expected trajectories inside the QD. Bottom: Temperature dependence of the amplitude of peaks 1 and 2.

Using the expression of the cyclotron radius $r_c = (h/eB)*(n_s/2\pi)^{1/2}$, and given the size of the cavity, the magnetic field position of the expected trajectories can easily be evaluated (Fig. 2). Two trajectories linking both point contacts correspond to maxima at 0.81T and 1.47T, and the minimum at 0.49T is associated with an electron trajectory reflecting electrons in the origin point contact. The uncertainty on the predicted peak positions comes from the experimental error on $n_s$ and the cavity diameter, and from the direction for the carrier injection. Following simple geometrical arguments, one could draw several other electron paths in the investigated magnetic field range, which should give rise to other minima and maxima in the magnetoconductance. However, all trajectories do not contribute equally to the magnetoconductance. As electron reflections on device boundaries are not always specular, trajectories involving a small number of such reflections have a larger contribution than more complicated paths.

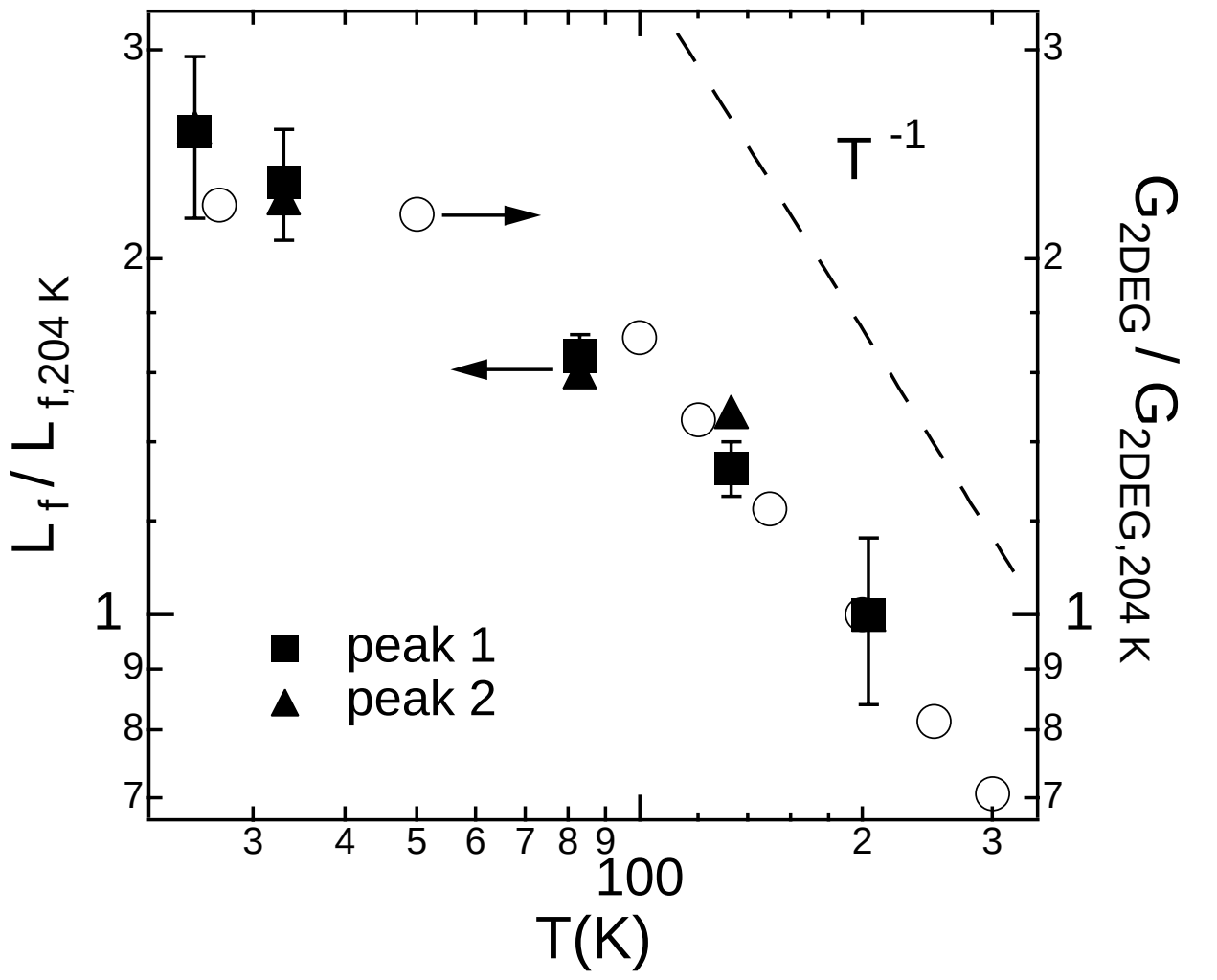

Fig. 3. Temperature dependence of the change in $L_f$ and in the longitudinal conductance of the 2DEG (both normalized to the value at 204 K). The dashed line represents a $T^{-1}$ law.

Focusing measurements [3] evidenced that the focusing peak amplitude $A_p$ depends on the trajectory length $L_e$ and the focusing mean free path $L_f$ according to the following expression :

$$A_p = A_0 \exp(-L_e/L_f), \qquad (1)$$

where $A_0$ is a constant. The maxima at $B$=0.81T (peak 1) and 1.47T (peak 2) correspond respectively to trajectory lengths $L_e$ ~ 596 nm and 700 nm, both larger than the electron mean free path, even at low temperature. However, ballistic effects can still be observed when $L_e$ is larger than $l_\mu$, as evidenced by Hirayama and Tarucha [4]. The peak amplitude $A_p$, shown on Fig. 2 as a function of the temperature, is obtained after substracting from each value of $G(B)$ the value of the conductance linearly interpolated between each side of the peak. As this method adds the amplitude of a maximum to that of the adjacent minima, it only gives an indication on the temperature dependence of $A_p$ and $L_f$, but not on their absolute value. Note also that the extraction of $A_p$ was not performed below 20 K, because UCFs amplitude becomes comparable to the amplitude of the focusing fluctuations.

In order to calculate $L_f$ from equ. (1), the value of $A_0$ has to be determined. This requires a reference value for $L_f$ at a given temperature, which is not available. Therefore, we only calculate the change in $L_f$ with respect to its value at 204K, $L_f/L_{f,204K}$. Fig. 3 compares these data to the relative change in the longitudinal conductance of the unpatterned two-dimensional electron gas, $G_{2DEG}$, with respect to the conductance at 204K, $G_{2DEG,204K}$. $G_{2DEG}/G_{2DEG,204K}$ is directly proportional to the elastic mean free path $l_\mu$. First, one sees that $G_{2DEG}/G_{2DEG,204K}$ saturates below ~50K, while $L_f/L_{f,204K}$ increases continuously as $T$ decreases. This difference may come from a different sensitivity of $l_\mu$ and $L_f$ to small angle elastic scattering mechanisms, as already evidenced by Heremans and co-workers in different types of experiments [3,5]. Despite this small discrepancy, the temperature dependence and the absolute values of both $G_{2DEG}/G_{2DEG,204K}$ and $L_f/L_{f,204K}$ are similar.

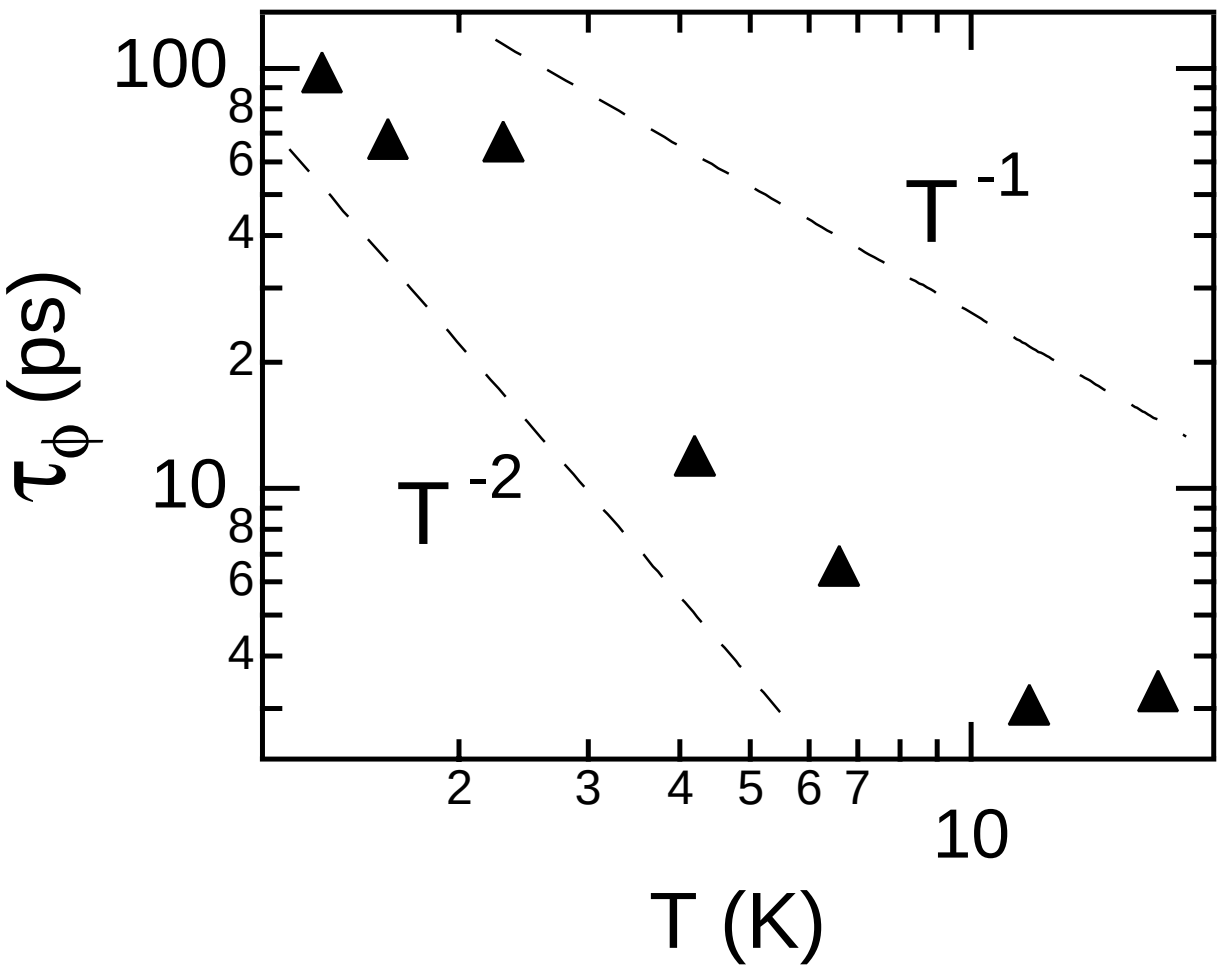

Fig. 4. Temperature dependence of the phase coherence time inside the cavity.
Dashed lines indicate $T^1$ and $T^2$ law.

Then, we focus on the phase coherence time $\tau_\phi$. We use the results of the random matrix theory applied to chaotic ballistic dots [6] to extract $\tau_\phi$ from the variance of UCFs [7]. The variance of the fluctuations is evaluated on high-pass filtered *G(B)* traces in order to remove the background contribution. As shown on Fig. 4, we find a temperature dependence of the form $\tau_\phi \propto T^{-b}$, where $1<b<2$, comparable to previous results in GaAs heterostructures [2]. This temperature dependence is qualitatively consistent with the 2D model of decoherence by electron-electron interactions in a diffusive system [2,8]. However, the absolute value of $\tau_\phi$ is larger than in ref. [2] by a factor of 2 to 4. This enhancement is likely to be due to the higher electron concentration of our substrate which plays an important role in electron-electron interaction effects.

## 4. Conclusion

In summary, we investigate the magnetoconductance of an InGaAs quantum dot. Our sample exhibits wide magnetoconductance fluctuations up to 204 K, caused by ballistic trajectories inside the dot. We calculate the temperature dependence of the focusing mean free path $L_f$. We find that $L_f$ follows approximately the same temperature dependence as that of the mobility mean free path, although $L_f$ does not saturate at low temperature. We calculate the electron phase coherence time up to 18 K, and find a roughly comparable temperature dependence as in GaAs heterostructure quantum dots, but with a larger absolute value by a factor of 2 to 4.

B.H., S.F. and C.G. acknowledge financial support from the F.R.I.A. This work was also supported by the Interuniversity attraction Pole (PAI-IUAP) program of the Belgian Government and by the Action de Recherche Concertee of the Communaute Francaise de Belgique .